\documentclass[twocolumn,showpacs,preprintnumbers,pra,hyperref]{revtex4}
\usepackage{amssymb}
\usepackage{amsfonts}
\usepackage{amsmath}
\usepackage{graphicx}

\begin{document}

\title{Variable frequency photonic crystals}
\author{Xiang-Yao Wu$^{a}$\footnote{E-mail: wuxy2066@163.com}, Ji Ma$^{a}$, Xiao-Jing Liu$^{a}$
\\ Jing-Hai Yang$^{a}$, Hong Li$^{a}$ and
 Wan-Jin Chen$^{a}$\footnote{E-mail: chenwwjj@126.com}}
\affiliation{$^{a}$ {\small Institute of Physics, Jilin Normal
University, Siping 136000 China}}

\begin{abstract}

In this paper, we have firstly proposed a new one-dimensional
variable frequency photonic crystals (VFPCs), and calculated the
transmissivity and the electronic field distribution of VFPCs with
and without defect layer, and considered the effect of defect
layer and variable frequency function on the transmissivity and
the electronic field distribution. We have obtained some new
characteristics for the VFPCs, which should be help to design a
new type optical devices.
\end{abstract}
\maketitle

 \vskip 8pt
 {\bf 1. Introduction} \vskip 8pt
Photonic crystals are artificial materials with periodic
variations in refractive index that are designed to affect the
propagation of light, which were first introduced theoretically by
Yablonovitch [1], and experimentally by John [2]. PCs, constructed
with periodic structure of artificial dielectrics or metallic
materials, have attracted many researchers in the past two decades
for their unique electromagnetic properties and scientific and
engineering applications [3-9]. These crystals indicate a range of
forbidden frequencies, called photonic band gap, as a result of
Bragg scattering of the electromagnetic waves passing through such
a periodical structure [10, 11]. As the periodicity of the
structure is broken by introducing a layer with different optical
properties, a localized defect mode will appear inside the band
gap. Enormous potential applications of PCs with defect layers in
different areas, such as light emitting diodes, filters and
fabrication of lasers have made such structures are interesting
research topic in this field.

Such materials are employed for the realization of diverse optical
devices, as for example distributed feedback laser [12, 13] and
optical switches [14, 15]. The addition of defects in the periodic
alternation, Or the realization of completely random sequences,
results in disordered photonic structures [16-18]. In the case of
one-dimensional disordered photonic structures, very interesting
physical phenomena Have been theoretically predicted or
experimentally observed.

In the paper, we have firstly proposed a new one-dimensional
variable frequency photonic crystals (VFPCs), which is made up of
variable frequency medium. The so-called variable frequency medium
is a new type optical medium, which can change the light frequency
when the light passes through the medium. For the conventional
medium, it does not change light frequency and change the light
wavelength. When light passes through a medium, it does not
exchange energy with light, the light frequencies should not be
changed, the medium is called conventional medium. When light
passes through a medium, it exchange energy with light, the light
frequencies should be changed, the medium is called variable
frequency medium. We can make the photonic crystals with the
variable frequency medium, which is called the variable frequency
photonic crystals (VFPCs). We should studied the transmissivity
and the electronic field distribution of the VFPCs and compare
them with the conventional PCs (CPCs), and obtained some new new
characteristics for the VFPCs, which should be help to design a
new type optical devices.

 \vskip 8pt
 {\bf 2. Transfer matrix, transmissivity and electronic field distribution of PCs}
\vskip 8pt

For one-dimensional conventional PCs, the calculations are
performed using the transfer matrix method [19], which is the most
effective technique to analyze the transmission properties of PCs.
For the medium layer $i$, the transfer matrices $M_i$ for $TE$
wave is given by [19]:
\begin{eqnarray}
M_{i}=\left(%
\begin{array}{cc}
 \cos\delta_{i} & -i\sin\delta_{i}/\eta_{i} \\
 -i\eta_{i}sin\delta_{i}
 & \cos\delta_{i}\\
\end{array}%
\right),
\end{eqnarray}
where $\delta_{i}=\frac{\omega}{c} n_{i} d_i cos\theta_i$, $c$ is
speed of light in vacuum, $\theta_i$ is the ray angle inside the
layer $i$ with refractive index $n_i=\sqrt{\varepsilon_i \mu_i}$,
$\eta_i=\sqrt{\varepsilon_i/\mu_i} cos\theta_i$,
$cos\theta_i=\sqrt{1-(n^2_0sin^2\theta_0/n^2_i)}$, in which $n_0$
is the refractive index of the environment wherein the incidence
wave tends to enter the structure, and $\theta_0$ is the incident
angle.

The final transfer matrix $M$ for an $N$ period structure is given
by:
\begin{eqnarray}
\left(%
\begin{array}{c}
  E_{1} \\
  H_{1} \\
\end{array}%
\right)&=&M_{B}M_{A}M_{B}M_{A}\cdot\cdot\cdot M_{B}M_{A}\left(%
\begin{array}{c}
  E_{N+1} \\
  H_{N+1} \\
\end{array}%
\right)
\nonumber\\&=&M\left(%
\begin{array}{c}
  E_{N+1} \\
  H_{N+1} \\
\end{array}%
\right)=\left(%
\begin{array}{c c}
  A &  B \\
 C &  D \\
\end{array}%
\right)
 \left(%
\begin{array}{c}
  E_{N+1} \\
  H_{N+1} \\
\end{array}%
\right),
\end{eqnarray}
where
\begin{eqnarray}
M=\left(%
\begin{array}{c c}
  A &  B \\
 C &  D \\
\end{array}%
\right),
\end{eqnarray}
with the total transfer matrix $M$, we can obtain the transmission
coefficient $t$, and the transmissivity $T$, they are
\begin{eqnarray}
t=\frac{E_{N+1}}{E_{1}}=\frac{2\eta_{0}}{A\eta_{0}+B\eta_{0}\eta_{N+1}+C+D\eta_{N+1}},
\end{eqnarray}
\begin{eqnarray}
T=t\cdot t^{*}.
\end{eqnarray}
Where $\eta_{0}=\eta_{N+1}=\sqrt{\frac{\varepsilon_0}{\mu_0}}$.

The electronic field distribution at position $x$ is [19]
\begin{eqnarray}
\left(%
\begin{array}{c}
  E(x) \\
  H(x) \\
\end{array}%
\right)&=&M_{A}(a-x)M_{B}(M_{A}M_{B})^{N-1}\left(%
\begin{array}{c}
  E_{N+1} \\
  H_{N+1} \\
\end{array}%
\right) \nonumber\\&=&\left(%
\begin{array}{cc}
 A'(x) & B'(x) \\
 C'(x) & D'(x)\\
\end{array}%
\right) \left(%
\begin{array}{c}
  E_{N+1} \\
  H_{N+1} \\
\end{array}%
\right),
\end{eqnarray}
with $E_{N+1}=E_{1}\cdot t$ and
$H_{N+1}=\sqrt{\varepsilon_0/\mu_0} \cdot E_{N+1}$, we have
\begin{eqnarray}
E(x)=(A'(x)+B'(x)\sqrt{\varepsilon_0/\mu_0})E_{1}\cdot t,
\end{eqnarray}
and
\begin{eqnarray}
|\frac{E(x)}{E_{1}}|^2= |A'(x)+B'(x)\sqrt{\varepsilon_0/\mu_0}|^2
\cdot|t|^2.
\end{eqnarray}

\begin{figure}[tbp]
\includegraphics[width=9 cm]{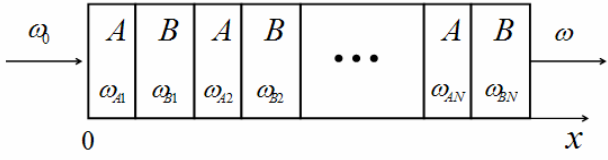}
\caption{one-dimensional variable frequency photonic crystals.}
\end{figure}

\begin{figure}[tbp]
\includegraphics[width=9 cm]{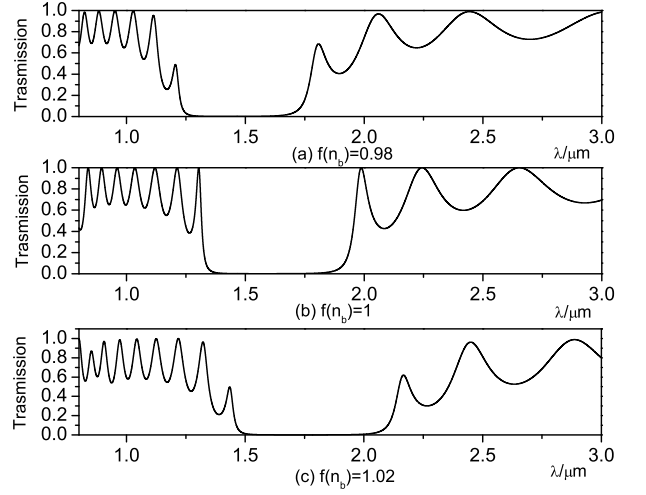}
\caption{Comparing the transmissivity of VFPCs with conventional
PCs. (a) VFPCs $f(n_b)=0.98$, (b) conventional PCs $f(n_b)=1.0$,
(c) VFPCs $f(n_b)=1.02$.}
\end{figure}

\begin{figure}[tbp]
\includegraphics[width=9 cm]{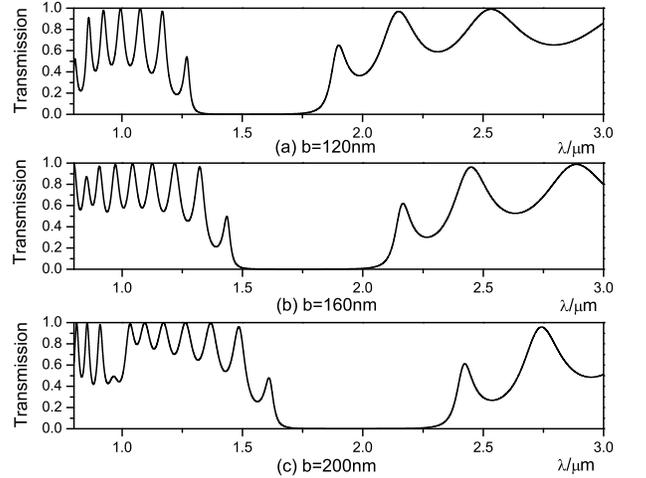}
\caption{The effect of the variable frequency medium thickness $b$
on the VFPCs transmissivity. (a) $b=120nm$, (b) $b=160nm$, (c)
$b=200nm$.}
\end{figure}

\vskip 8pt {\bf 3. Transfer matrix and transmissivity of the
VFPCs}\vskip 8pt

We have firstly proposed a new one-dimensional photonic crystals
constituting by variable frequency medium. When light passes
through the variable frequency medium, the light frequency can be
written as
\begin{eqnarray}
\omega=f(n)\omega_0,
\end{eqnarray}
where $\omega_0$ is the incident light frequency, $\omega$ is the
frequency of light in the variable frequency medium, $n$ is the
refractive index of variable frequency medium, and $f(n)$ is
called variable frequency function, which expresses the degree
changing the light frequency for the variable frequency medium.
When the variable frequency function $f(n)>1$, the light frequency
should be increased in variable frequency medium, when $f(n)<1$,
the light frequency should be decreased. When $f(n_b)=1$ the
variable frequency medium become the conventional medium.

For the VFPCs, its transfer matrices are similar to the
conventional transfer matrices (1), they are
\begin{eqnarray}
M_{A_i}=\left(%
\begin{array}{cc}
 \cos\delta_{a_i} & -i\sin\delta_{a_i}/\eta_{a} \\
 -i\eta_{a}sin\delta_{a_i}
 & \cos\delta_{a_i}\\
\end{array}%
\right)(i=1, 2, \cdot\cdot\cdot, N),
\end{eqnarray}
\begin{eqnarray}
M_{B_i}=\left(%
\begin{array}{cc}
 \cos\delta_{b_i} & -i\sin\delta_{b_i}/\eta_{b} \\
 -i\eta_{b}sin\delta_{b_i}
 & \cos\delta_{b_i}\\
\end{array}%
\right)(i=1, 2, \cdot\cdot\cdot, N),
\end{eqnarray}
For the structure of one-dimensional VFPCs $(AB)^N$, where the
medium $B$ is the variable frequency medium, the medium $A$ is the
conventional medium, and $N$ is the period numbers. The light
frequencies of every medium are as follows:
\begin{eqnarray}
\omega_{A1}=\omega_{0},
\end{eqnarray}
\begin{eqnarray}
\omega_{B1}=f(n_b)\omega_{0},
\end{eqnarray}
\begin{eqnarray}
\omega_{A2}=\omega_{B1}=f(n_b)\omega_{0},
\end{eqnarray}
\begin{eqnarray}
\omega_{B2}=f(n_b)\omega_{A2}=f^{2}(n_b)\omega_{0},
\end{eqnarray}
\begin{eqnarray}
\omega_{Ai}=f^{i-1}(n_b)\omega_{0},
\end{eqnarray}
\begin{eqnarray}
\omega_{Bi}=f^{i}(n_b)\omega_{0},
\end{eqnarray}
which are shown in FIG. 1, and the corresponding phases are:
\begin{eqnarray}
\delta_{a_1}=\frac{\omega_0}{c}\cdot n_{a}\cdot a,
\end{eqnarray}
\begin{eqnarray}
\delta_{b_1}=\frac{f(n_b)\cdot \omega_0}{c}\cdot n_{b}\cdot b,
\end{eqnarray}
\begin{eqnarray}
\delta_{a_2}=\frac{f(n_b)\cdot \omega_0}{c}\cdot n_{a}\cdot a,
\end{eqnarray}
\begin{eqnarray}
\delta_{b_2}= \frac{f^2(n_b)\cdot \omega_0}{c}\cdot n_{b}\cdot b,
\end{eqnarray}
\begin{eqnarray}
\delta_{a_i}=\frac{f^{i-1}(n_b)\cdot \omega_0}{c}\cdot n_{a}\cdot
a,
\end{eqnarray}
\begin{eqnarray}
\delta_{b_i}= \frac{f^{i}(n_b)\cdot \omega_0}{c}\cdot n_{b}\cdot
b.
\end{eqnarray}

\begin{figure}[tbp]
\includegraphics[width=9 cm]{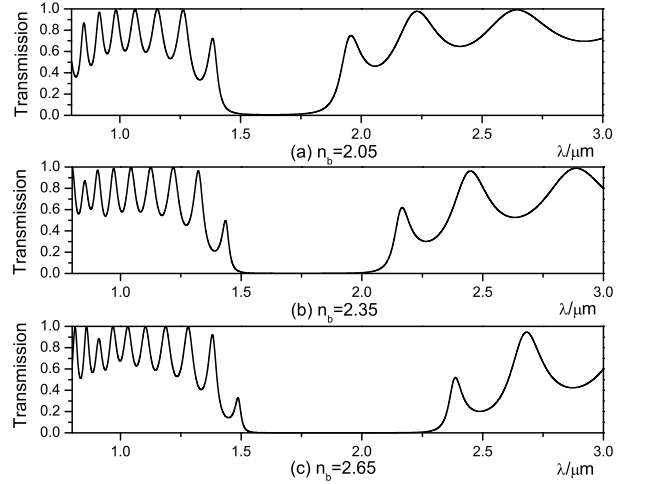}
\caption{The effect of the variable frequency medium refractive
index $n_b$ on the VFPCs transmissivity. (a) $n_b=2.05$, (b)
$n_b=2.35$, (c) $n_b=2.65$.}
\end{figure}

\begin{figure}[tbp]
\includegraphics[width=9 cm]{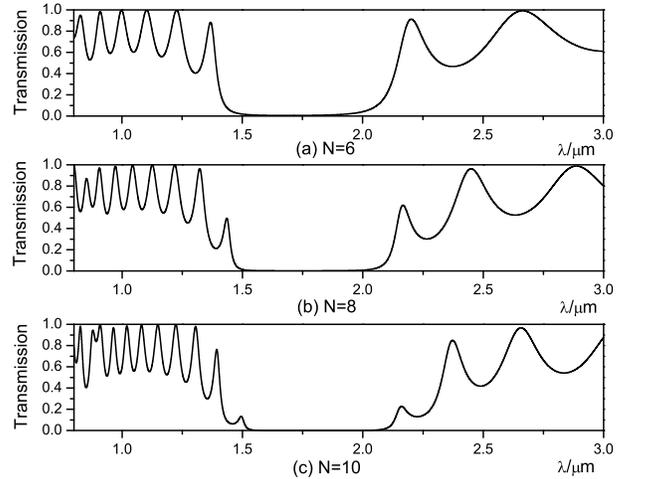}
\caption{The effect of the period number $N$ on the VFPCs
transmissivity. (a) $N=6$, (b) $N=8$, (c) $N=10$.}
\end{figure}
 \vskip 8pt
 {\bf 4. Numerical result}
 \vskip 8pt
In this section, we report our numerical results of VFPCs. The
VFPCs main parameters are: The medium $A$ is conventional medium,
its refractive indexes $n_a=1.38$, thickness $a=298nm$, the medium
$B$ is variable frequency medium, its refractive indexes
$n_b=2.35$, thickness $b=160nm$, and variable frequency function
$f(n_b)$, the defect layer medium $D$ refractive indexes
$n_d=2.97$, thickness $d=380nm$. The structure of VFPCs is
$(AB)^{8}$. In FIG. 2, we study the effect of the variable
frequency function $f(n_b)$ on VFPCs transmissivity, the variable
frequency function $f(n_b)$ in the FIG. 2 (a), (b) and (c) are
$0.98$, $1$ and $1.02$, respectively. The FIG. 2 (b) is the
transmissivity of conventional PCs because of $f(n_b)=1$, and FIG.
2 (a) and (c) are the transmissivity of VFPCs because of $f\neq
1$. Comparing the transmissivity of VFPCs with conventional PCs,
we can obtain the new results: (1) When the variable frequency
function $f(n_b)<1$ (FIG. 2 (a)), the band gaps blue shift, and
the band gaps width decrease. (2) When the variable frequency
function $f(n_b)>1$ (FIG. 2 (c)), the band gaps red shift, and the
band gaps width increase. From the results (1) and (2), we find
the variable frequency function is an important factor of effect
on transmissivity.
\begin{figure}[tbp]
\includegraphics[width=9 cm]{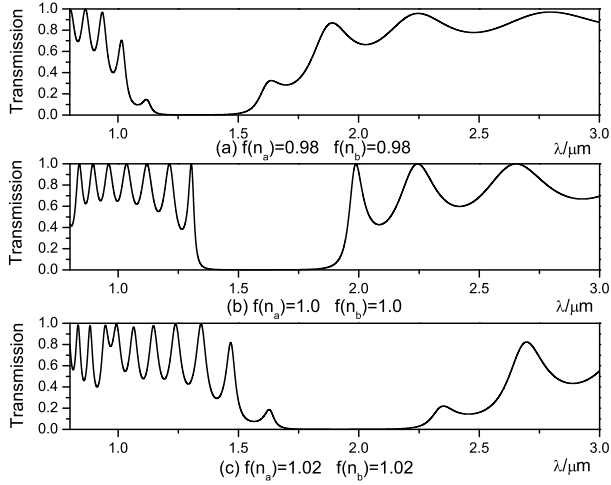}
\caption{The effect of the variable frequency functions $f(n_a)$
and $f(n_b)$ on the VFPCs transmissivity. (a) $f(n_a)=0.98$
$f(n_b)=0.98$, (b) $f(n_a)=1.0$ $f(n_b)=1.0$, (c) $f(n_a)=1.02$
$f(n_b)=1.02$.}
\end{figure}

\begin{figure}[tbp]
\includegraphics[width=9 cm]{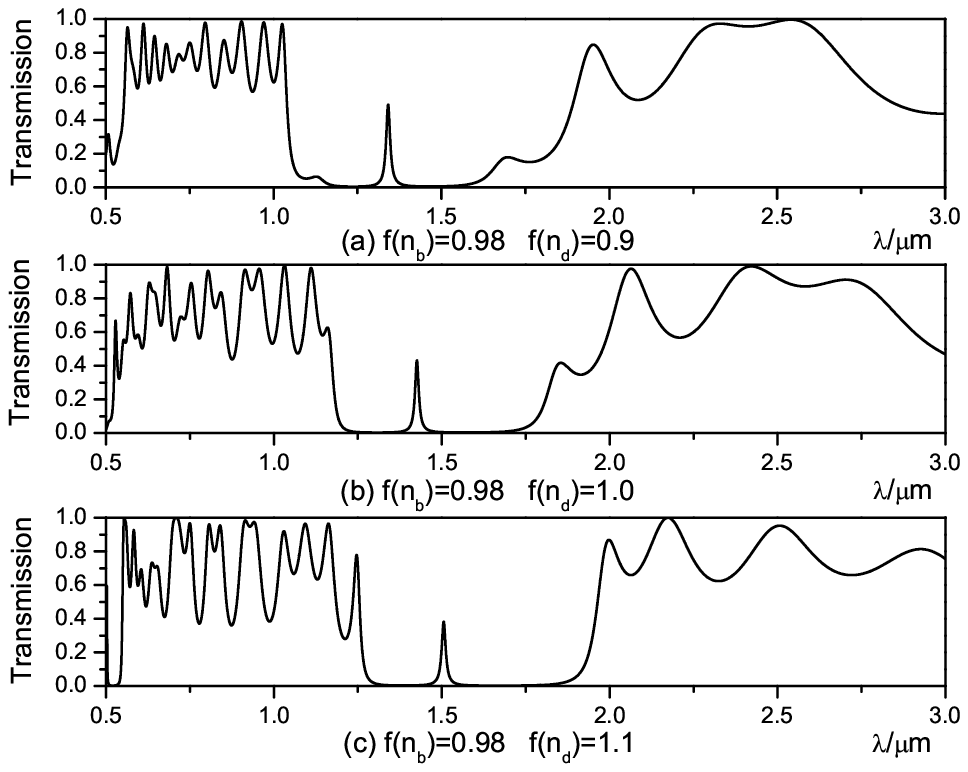}
\caption{The effect of the defect layer variable frequency
function $f(n_d)$ on the structure $(AB)^4D(AB)^4$ VFPCs
transmissivity. (a) $f(n_b)=0.98$, $f(n_d)=0.9$, (b)
$f(n_b)=0.98$, $f(n_d)=1.0$, (c) $f(n_b)=0.98$, $f(n_d)=1.1$.}
\end{figure}
\begin{figure}[tbp]
\includegraphics[width=9 cm]{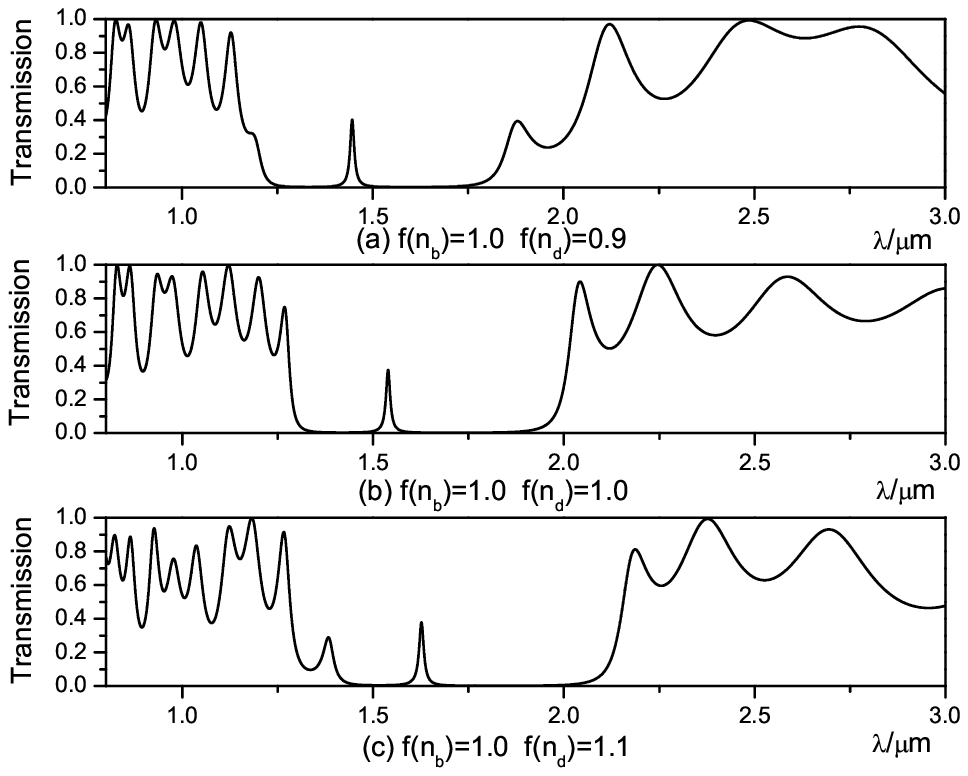}
\caption{The effect of the defect layer variable frequency
function $f(n_d)$ on the structure $(AB)^4D(AB)^4$ VFPCs
transmissivity. (a) $f(n_b)=1.0$, $f(n_d)=0.9$, (b) $f(n_b)=1.0$,
$f(n_d)=1.0$, (c) $f(n_b)=1.0$, $f(n_d)=1.1$.}
\end{figure}

\begin{figure}[tbp]
\includegraphics[width=9 cm]{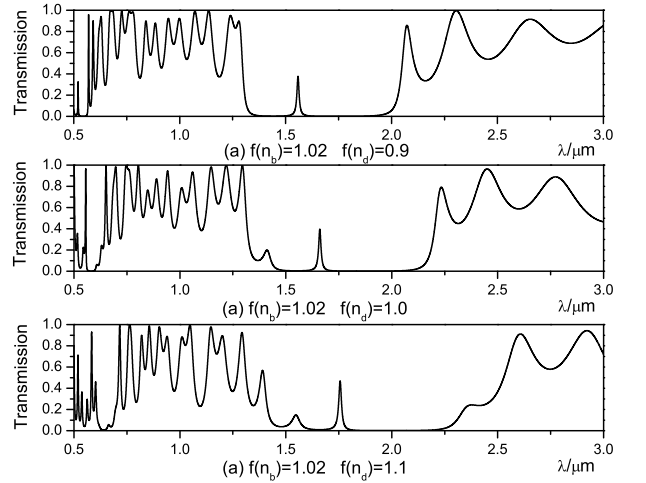}
\caption{The effect of the defect layer variable frequency
function $f(n_d)$ on the structure $(AB)^4D(AB)^4$ VFPCs
transmissivity. (a) $f(n_b)=1.02$, $f(n_d)=0.9$, (b)
$f(n_b)=1.02$, $f(n_d)=1.0$, (c) $f(n_b)=1.02$, $f(n_d)=1.1$.}
\end{figure}

In FIG. 3, we study the effect of the variable frequency medium
thickness on the VFPCs transmissivity, wherein variable frequency
function $f(n_b)=1.02$. The FIG. 3 (a), (b) and (c) thickness $b$
are $120nm$, $160nm$ and $200nm$, respectively. From FIG. 3 (a),
(b) and (c), we can find when the variable frequency medium
thickness b increase the band gaps red shift, and the band gaps
width increase. In FIG. 4, we study the effect of the variable
frequency medium refractive index on the VFPCs transmissivity,
wherein variable frequency function $f(n_b)=1.02$. The FIG. 4 (a),
(b) and (c) refractive index $n_b$ are $2.05$, $2.35$ and $2.65$,
respectively. From FIG. 4 (a), (b) and (c), we can find when the
variable frequency medium refractive index $n_b$ increase the band
gaps red shift, and the band gaps width increase. In FIG. 5, we
study the effect of the period number $N$ on the VFPCs
transmissivity, wherein variable frequency function $f(n_b)=1.02$.
The FIG. 5 (a), (b) and (c) period number $N$ are $6$, $8$ and
$10$, respectively. From FIG. 5 (a), (b) and (c), we can find when
the VFPCs period numbers increase the band gaps width increase. In
the conventional PCs, the band gaps width are unchanged when the
period numbers change. In FIG. 6, we consider the media $A$ and
$B$ are all variable frequency medium, and study the effect of the
variable frequency functions $f(n_a)$ and $f(n_b)$ on VFPCs
transmissivity, the variable frequency functions $f(n_a)$ and
$f(n_b)$ in the FIG. 6 (a), (b) and (c) are all $0.98$, $1$ and
$1.02$, respectively. Comparing the transmissivity of VFPCs with
conventional PCs (FIG. 6 (b)), we can obtain the new results: (1)
When the variable frequency functions $f(n_a)<1$ and $f(n_b)<1$
(FIG. 6 (a)), the band gaps blue shift, and the band gaps width
decrease obviously. (2) When the variable frequency functions
$f(n_a)>1$ and $f(n_b)>1$ (FIG. 6 (c)), the band gaps red shift,
and the band gaps width increase obviously. In FIGs. 7, 8 and 9,
we should study the effect of defect layer on the transmissivity,
the structures of VFPCs are $(AB)^4D(AB)^4$, the variable
frequency function of medium $B$ $f(n_b)=0.98$, $f(n_b)=1.0$ and
$f(n_b)=1.02$ corresponding to FIGs. 7, 8 and 9, the medium $A$ is
conventional medium. For the defect layer medium $D$, its
thickness, refractive index are: $d=380nm$ and $n_d=2.97$, and the
variable frequency function $f({n_d})$ are $0.9$, $1.0$ and $1.1$
corresponding to the figures (a), (b) and (c) of FIGs. 7 to 9.
From FIGs. 7 to 9, we can find when the variable frequency
function of defect layer $f{n_d}$ increases, the defect model
position red shift and intensity decreases. In the FIG. 10, the
VFPCs structure is $(AB)^4D(AB)^4$, the variable frequency
functions of medium $B$ and defect layer medium $D$ are
$f(n_b)=0.98$ and $f(n_d)=1.1$, and the medium $A$ is conventional
medium. In the FIG. 10 (a), (b) and (c), the refractive indices of
defect layer medium $D$ are $n_d=2.97+i0.02$ (absorbing medium),
$n_d=2.97$ (conventional medium) and $n_d=2.97-i0.02$ (active
medium), respectively. Comparing with the conventional medium
defect layer (FIG. 10 (b)), we can find the absorbing medium
defect layer (FIG. 10 (a)) make the defect model intensity
decrease and the active medium defect layer (FIG. 10 (c)) make the
defect model intensity increase. In FIG. 11, we calculate the
electronic field distribution of the VFPCs structure $(AB)^8$, and
consider the influence of variable frequency function $f(n_b)$ on
VFPCs electronic field distribution. The variable frequency
function $f(n_b)=0.97$, $1$ and $1.03$ are corresponding to the
dash dot line, solid line and dot line of electronic field
distribution, respectively. comparing with the conventional PCs
($f(n_b)=1$), we can obtain new results: (1) When the variable
frequency function $f(n_b)< 1$ the peak values of VFPCs electronic
field distribution decreased, the distribution curve left shift.
(2) When the variable frequency function $f(n_b)> 1$ the peak
values of VFPCs electronic field distribution increased, the
distribution curve right shift. In FIGs. 12 and 13, the VFPCs
structure is $(AB)^4D(AB)^4$, the medium $A$ and defect layer
medium $D$ are conventional media and medium $B$ is the variable
frequency medium. We consider the effect of the defect layer on
electronic field distribution of VFPCs. In FIG. 12, the variable
frequency function of medium $B$ $f(n_b)=0.98$ and the defect
layer medium $D$ refractive index $n_d=2.59$, thickness $d=200nm$.
In FIG. 13, the variable frequency function of medium $B$
$f(n_b)=1.02$ and the defect layer medium $D$ refractive index
$n_d=2.59$, thickness $d=88nm$. Comparing FIGs. 12 and 13 with
FIG. 11, we can find the electronic field distribution has been
local enhanced in the vicinity of defect layer.

\begin{figure}[tbp]
\includegraphics[width=9 cm]{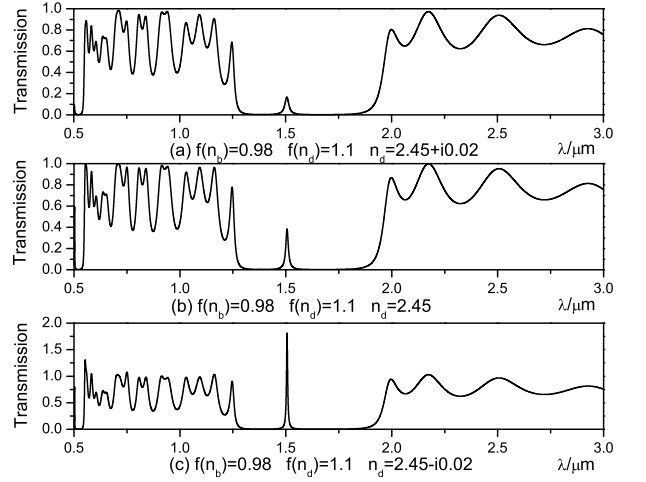}
\caption{The effect of the defect layer refractive index on the
structure $(AB)^4D(AB)^4$ VFPCs transmissivity. (a)
$n_d=2.59+i0.02$, (b) $n_d=2.59$, (c) $n_d=2.59-i0.02$.}
\end{figure}

\begin{figure}[tbp]
\includegraphics[width=9 cm]{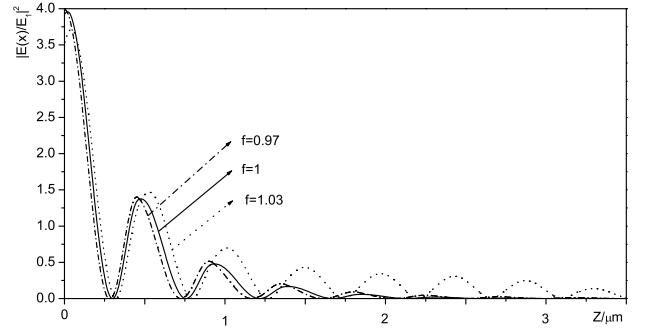}
\caption{The effect of the variable frequency functions $f(n_b)$
on the electronic field distribution $|E(x)/E_1|^2$ for the
structure $(AB)^8$ VFPCs. (a) $f(n_b)=0.97$ dash dot line, (b)
$f(n_b)=1$ solid line, (c) $f(n_b)=1.03$ dot line.}
\end{figure}

\begin{figure}[tbp]
\includegraphics[width=9 cm]{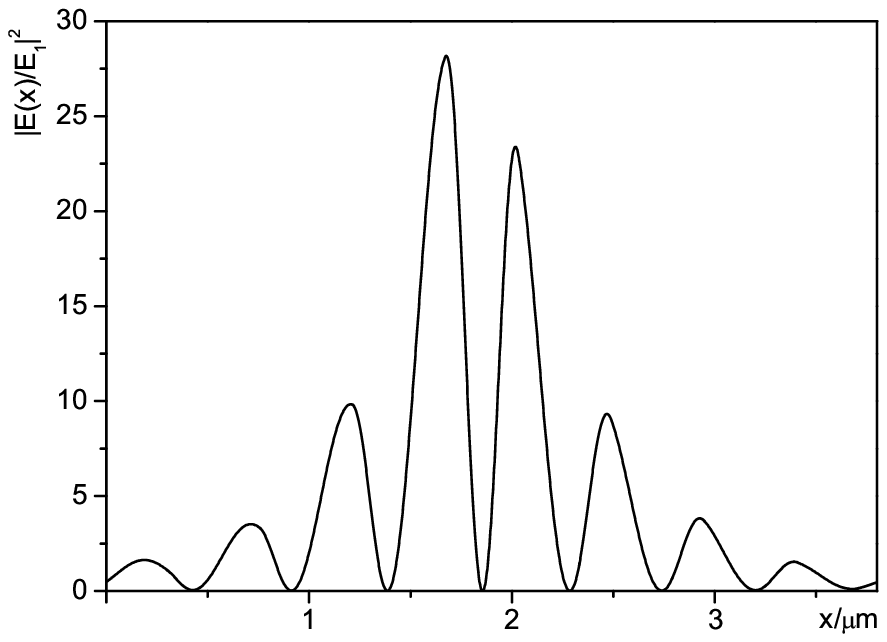}
\caption{The effect of the defect layer $D$ on the electronic
field distribution $|E(x)/E_1|^2$ for the structure
$(AB)^4D(AB)^4$ VFPCs. The $f(n_b)=0.98$ and $d=200nm$.}
\end{figure}

\begin{figure}[tbp]
\includegraphics[width=9 cm]{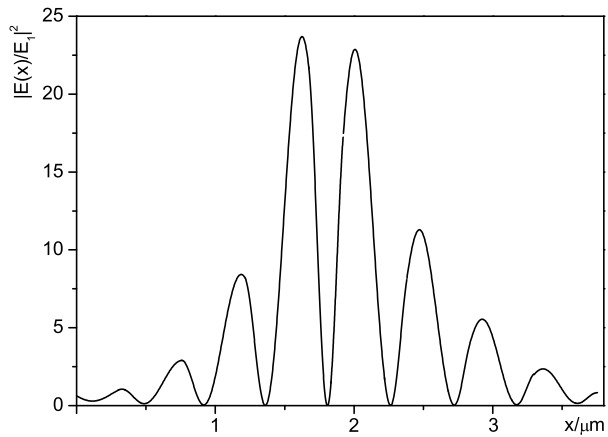}
\caption{The effect of the defect layer $D$ on the electronic
field distribution $|E(x)/E_1|^2$ for the structure
$(AB)^4D(AB)^4$ VFPCs. The $f(n_b)=1.02$ and $d=88nm$.}
\end{figure}

 \vskip 8pt
{\bf 5. Conclusion}
 \vskip 8pt
In summary, we have studied the transmissivity and the electronic
field distribution of one-dimensional VFPCs with and without
defect layer, and compare them with the conventional PCs. We
obtained some new results: (1) When the variable frequency
function $f(n_b)<1$, the band gaps blue shift, and the band gaps
width decrease. (2) When the variable frequency function
$f(n_b)>1$, the band gaps red shift, and the band gaps width
increase, i.e., the variable frequency function is an important
factor of effect on transmissivity. (3) When the VFPCs period
numbers increase the band gaps width increase. In the conventional
PCs, the band gaps are unchanged when the period numbers increase.
(4) When there is defect layer, we can find when the variable
frequency function of defect layer $f{n_d}$ increases, the
position of defect model red shift and its intensity decreases.
(5) Comparing with the conventional medium, we can find the
absorbing medium make the defect model intensity decrease and the
active medium make the defect model intensity increase. (6) When
the variable frequency function $f(n_b)< 1$ the peak values of
VFPCs electronic field distribution decreased, the distribution
curve left shift. (7) When the variable frequency function
$f(n_b)> 1$ the peak values of VFPCs electronic field distribution
increased, the distribution curve right shift. (8) When there is
defect layer, we can find the electronic field distribution has
been local enhanced in the vicinity of defect layer. These new
characteristics of VFPCs should be help to design a new type
optical devices.

\vskip 8pt {\bf 6.  Acknowledgment} \vskip 8pt

This work is supported by Scientific and Technological Development
Foundation of Jilin Province, Grant Number: 20130101031JC.

\end{document}